\documentclass[letterpaper, 10 pt, conference]{ieeeconf}  %

\IEEEoverridecommandlockouts                              %
\usepackage{palatino,epsfig,amssymb,cite,color}
\usepackage[fleqn]{amsmath}

\newtheorem{prop}{Proposition}
\newtheorem{lemma}{Lemma} 
\newtheorem{theorem}{Theorem}

\def\QED{~\rule[-1pt]{5pt}{5pt}\par}

\DeclareMathAlphabet{\matheur}{U}{eur}{m}{n}
\DeclareMathAlphabet{\matheurb}{U}{eur}{b}{n}
\DeclareMathAlphabet{\matheus}{U}{eus}{m}{n}
\DeclareMathAlphabet{\matheuf}{U}{euf}{m}{n}

\newcommand{\hs}{\hspace{4mm}}

\renewcommand{\t}{^{\mbox{\tiny\sf T}}}
\newcommand{\IC}{\mathbb{C}}
\newcommand{\IR}{\mathbb{R}}

\newcommand{\IS}{\mathbb{S}}

\newcommand{\sig}{{\sigma}}

\newcommand{\eps}{\varepsilon}
\newcommand{\delo}{\delta_p}

\newenvironment{mat}{\left[\begin{array}}{\end{array}\right]}

\DeclareMathAlphabet{\matheur}{U}{eur}{m}{n}
\DeclareMathAlphabet{\matheurb}{U}{eur}{b}{n}
\DeclareMathAlphabet{\matheus}{U}{eus}{m}{n}
\DeclareMathAlphabet{\matheuf}{U}{euf}{m}{n}

\newcommand{\rhol}{{\varrho}}

\newcommand{\RHinf}{{\bf RH}_\infty}

\usepackage{fancyhdr, url}

\fancypagestyle{firstpage}{\lfoot{\scriptsize \copyright 2020 IEEE. Personal use of this material is permitted. Permission from IEEE must be obtained for all other uses, in any current or future media, including reprinting/republishing this material for advertising or promotional purposes, creating new collective works, for resale or redistribution to servers or lists, or reuse of any copyrighted component of this work in other works. The final version of record is available at \url{https://doi.org/10.1109/CDC42340.2020.9304320}} \cfoot{}}

\begin{document}

\title{\LARGE \bf
Robust Instability Analysis \\ with Application to Neuronal Dynamics}
\author{Shinji Hara, Tetsuya Iwasaki*, Yutaka Hori%
\thanks{
This work was supported in part by the Ministry of Education, Culture, Sports, Science and Technology in Japan 
through Grant-in-Aid for Scientific Research (A) 21246067 and (B) 18H01464. 
S.~Hara is with 
Tokyo Institute of Technology, 
{\small Shinji\_Hara@ipc.i.u-tokyo.ac.jp}, 
~T.~Iwasaki is with 
 University of California Los Angels, 
{\small tiwasaki@ucla.edu}, 
~Y.~Hori is with 
Keio University, 
{\small yhori@appi.keio.ac.jp}. 
}
}

\maketitle
\thispagestyle{empty}
\thispagestyle{firstpage}
\begin{abstract}
This paper is concerned with robust instability analysis of linear 
feedback systems subject to a dynamic uncertainty. The work is motivated by, 
and provides a basic foundation for, a more challenging problem of analyzing  
persistence of oscillations in nonlinear dynamical systems.
We first formalize the problem for SISO  LTI systems by introducing a notion of 
the robust instability radius (RIR). 
We provide a method for calculating the RIR exactly for a certain 
class of systems and show that it works well for a class of second order systems. 
This result is applied to the FitzHugh-Nagumo model for neuronal dynamics, 
and the effectiveness is confirmed by numerical simulations, where 
we properly care for the change of the equilibrium point. 
\end{abstract}
\section{Introduction}
\label{sec:Intro}

There are a number of oscillatory phenomena that play functional roles in biology. 
For scientific understanding as well as for engineering applications, it is desired to 
characterize robustness of such oscillations against perturbations. In general, however, 
existence and stability of limit cycle oscillations are difficult to analyze exactly, 
even for the nominal (unperturbed) case. A practical approach to this challenging
problem is to focus on the robust instability property of an equilibrium point, which 
would not rigorously guarantee, but yield in many cases, oscillations around the
equilibrium sustained under perturbations. 

The classical bifurcation analysis is in line with this approach and examines how stability 
properties of an equilibrium point changes under a parametric perturbation
(see e.g. \cite{Kim:IEESB2006, Waldherr:Automatica2011} for numerical methods 
from the view point of robust instability). 
However, realistic perturbations often possess dynamics, and hence an extension to 
the so-called robust bifurcation analysis \cite{Inoue:ECC2013} is desired. Apart from the 
major issue associated with the fact that the equilibrium point can move due to 
perturbations \cite{Inoue:ECC2014}, the problem essentially reduces to a robust instability 
analysis of linear systems through the Hartman-Grobman theorem. 

In contrast with robust stability analysis, robust instability analysis is not easy %
because instability may be sustained even when some of the nominally unstable poles 
become stable due to a certain perturbation. 
In other words, we need to keep track of the behavior of all nominally unstable modes.  
Consequently, the traditional small-gain type argument based on the $\infty$-norm may not work in general, 
meaning that the $L_\infty$-norm condition provides only sufficient conditions 
for the robust instability as seen in e.g., \cite{Inoue:ECC2013, Inoue:ECC2014}. 
Moreover, such conditions could be very conservative.

There is another interesting point on the issue.
The robust instability problem is equivalent to strong stabilization by a minimum-norm, stable controller. 
The condition for strong stabilizability has been known, 
but the order of a strongly stabilizing controller 
is unknown \cite{Youla:Automatica1974}. 
Minimization of a norm on some closed-loop transfer functions has been considered in the literature, 
but only partial solutions have been obtained due to the difficulty in enforcing the stability 
constraint on the controller, e.g. \cite{Zeren:Automatica2000}. 
Therefore, it is required to develop a new fundamental theory for robust instability analysis.  

In this paper, we make a first step toward development of a general theory for robust 
nonlinear oscillations. Specifically, we focus on linear systems, examine the origins
of the difficulties associated with the robust instability analysis, and propose an idea for 
exact characterization of the robust instability property. To this end, we consider a class 
of uncertain feedback systems consisting of unstable nominal part $g(s)$ and stable perturbation 
$\delta(s)$, both of which are single-input-single-output (SISO) linear time-invariant (LTI) systems. 
We first formalize the problem by defining a notion of the robust instability radius (RIR) 
as the maximum allowable $\infty$-norm of $\delta(s)$ that maintains instability of the feedback system. 
After showing lower and upper bounds of RIR, we present numerical examples of second order systems 
to illustrate when the bounds can be conservative/tight, helping us 
understand the reason why computing the exact RIR is hard in general. 
We then provide an idea for showing tightness of the lower bound and thus finding the 
exact RIR, using a first order all-pass function as a stabilizing perturbation. 
It is proven that the idea works well for a class of second order systems. 
This theoretical result is applied to the FitzHugh-Nagumo model \cite{FHNmodel}
for neuronal dynamics, and its effectiveness is confirmed by numerical 
simulations, where we properly care for the change of the equilibrium point. 

{\bf Notation:}  
$\Re(s)$ denotes the real part of a complex number $s$. 
The open left (resp. right) half complex plane is abbreviated as OLHP (resp. ORHP).
The set of proper stable real rational functions is denoted by $\RHinf$. 
For a transfer function $g(s)$, the $H_\infty$ norm and $L_\infty$ norm are defined by 
$\|g\|_{H_\infty} := \sup_{s \in \mbox{\tiny ORHP}} |g(s)|$ and
$\|g\|_{L_\infty} := \sup_{\omega \in \IR} |g(j\omega)|$, respectively.

\section{Robust Instability Radius (RIR)}
\label{subsec:RIR}

Given an unstable unity feedback system with open-loop transfer function 
$h(s)$, we analyze the robust instability property against a general class of perturbations. 
To explain, let us take a typical case of multiplicative uncertainties, {\it i.e.}, 
the perturbed loop transfer function is given by  
$\tilde{h}(s) = (1 + \delta(s))h(s)$, where $\delta(s)$ is an uncertain $\RHinf$ function. 
Similarly to the robust stability analysis,  
the corresponding characteristic equation is given by  
\begin{equation} \label{eq:che}
 1 - \delta(s) g(s) = 0,  
\end{equation}  
where 
we assume positive feedback, and 
$g(s) := h(s)/(1 - h(s))$ is called the complementary sensitivity 
function. In this paper, we consider (\ref{eq:che}) as the basic 
characteristic equation with an arbitrary unstable $g(s)$, not restricted to 
the complementary sensitivity function. 
For technical reasons, we assume that $g(s)$ has no pole on the imaginary axis. 

We are interested in determining how much perturbation is allowed before the closed-loop system becomes stable. 
That is, the objective is to find the smallest norm stable perturbation $\delta(s)$ such that 
(\ref{eq:che}) has all roots in the OLHP, 
where no unstable pole/zero cancellations are allowed between $g(s)$ and $\delta(s)$.

The robust instability radius (RIR), denoted by $\rho_*\in\IR$, with respect to real rational dynamic perturbation 
$\delta\in\RHinf$, is defined as the smallest magnitude of the perturbation that internally stabilizes the system: 
\begin{equation} \label{rho}
\rho_*:=\inf_{\delta\in\IS(g)}~\|\delta\|_{H_\infty},
\end{equation}
where $\IS(g)$ is the set of real-rational, proper, stable transfer functions 
internally stabilizing $g(s)$, {\it i.e.}, 
\begin{equation} \label{IS}
\begin{array}{r}
\hspace{-1mm}
\IS(g) := \{ \delta\in\RHinf:~ \delta(s)g(s)=1 ~ \Rightarrow ~ \Re(s)<0, ~~\\ 
\delta(s)=0, ~ \Re(s)>0 ~ \Rightarrow ~ |g(s)| < \infty ~ \}
\end{array}
\end{equation}
When the perturbation is parametric, the real and complex RIRs are defined by
\begin{equation} \label{rhostatic}
\rho_r:=\inf_{\delta\in\IS_r(g)}~|\delta|, \hs
\rho_c:=\inf_{\delta\in\IS_c(g)}~|\delta|,
\end{equation}
where $\IS_r(g)\subset\IR$ and $\IS_c(g)\subset\IC$ are defined as in (\ref{IS}) 
by replacing $\RHinf$ with $\IR$ and $\IC$, respectively. 
It is clear by definition that $\rho_*\leq\rho_r$ and $\rho_c\leq\rho_r$.
However, it turns out that the relationship between $\rho_*$ and $\rho_c$ depends on $g(s)$ 
as we will see later by numerical examples.
This is in stark contrast with the robust stability analysis in which the robust stability radius with respect to 
the dynamic uncertainty coincides with that for the complex parametric uncertainty. 
The fact that $\rho_*\neq\rho_c$ makes the robust instability analysis for dynamic uncertainty extremely difficult.

The RIRs for the parametric uncertainty case can readily be calculated.
For example, the stability region $\IS_c(g)$ can be fully characterized within the framework of 
the generalized frequency variable \cite{HTI:TAC2014},
and it is easy to compute $\rho_r$ and $\rho_c$ by gridding the frequency at least. 

In contrast, the RIR for the dynamic uncertainty case is not easy to compute. 
It is noticed that our problem of robust instability has a clear correspondence with the so called 
strong stabilization, {\it i.e.}, $g(s)$ and $\delta(s)$ respectively correspond to the unstable plant 
and the stabilizing controller, which is required to be stable. 
Let us recall a classical result on strong stabilization by Youla 
{\em et al.} \cite{Youla:Automatica1974}: 
{\em
The robust instability radius $\rho_*$ for an unstable transfer function $g(s)$ is finite 
if and only if the Parity Interlacing Property (PIP) is satisfied, {\it i.e.},  
the number of unstable real poles of $g(s)$ between any pair of real zeros 
in the closed right half complex plane (including zero at $\infty$) is even.
}

\section{Preliminaries}
\label{sec:Preliminaries}
\subsection{Lower and Upper Bounds of RIR}
\label{subsec:LUbounds}

We here provide several results on lower and upper bounds of RIR as a preliminary.
First note by the definition of $\rho_*$ that there exists $\delta(s)$ such that $\|\delta\|_{H_\infty}=\rho_*$ 
and all the roots of $1 - \delta(s)g(s)=0$ satisfy $\Re(s)\leq0$ with at least one $s$ on 
the imaginary axis, say $s=j\omega_c$, because otherwise it is impossible to stabilize $g(s)$
 by stable $\delta(s)$ of norm arbitrarily close to $\rho_*$ due to continuity of the characteristic roots. 
Thus the critical perturbation $\delta(s)$ has to satisfy 
\begin{equation} \label{cp}
\delta(j\omega_c) = 1/g(j\omega_c), \hs |\delta(j\omega_c)|\leq\rho_*.
\end{equation}
Noting that $|g(j\omega_c)|\leq\|g\|_{L_\infty}$, a lower bound on $\rho_*$ can be obtained 
in terms of the $L_\infty$ norm of $g(s)$, which was noted in \cite{Inoue:ECC2013}, as follows. 

\smallskip

\begin{prop} \label{prop:lbub} 
{\em 
\begin{equation} \label{lub}
\vspace*{-2mm}
 \rhol_p \leq \rho_* \leq \rho_r , \hs \rhol_p:=1/\|g\|_{L_\infty} .
\end{equation}
} 
\end{prop}

\smallskip  

We can readily see that  $\rhol_p$ is also a lower bound on $\rho_c$, {\it i.e.}, $\rho_c \geq \rhol_p$ , 
since $\rho_c=1/|g(j\omega)|$ for some $\omega$ as explained above. 
Graphically, we see that $\rhol_p=\rho_c$ holds when the projection of
the origin onto the Nyquist plot of $1/g(s)$ is on the boundary of $\IS_c(g)$.

We now provide special cases where we may possibly get a better 
lower bound on $\rho_*$ than (\ref{lub}) or the exact value of $\rho_*$, 
where we evaluate the static gain of $g(s)$.  

\smallskip 

\begin{prop} \label{prop:lb0} 
{\em 
Given a real-rational, strictly proper transfer function $g(s)$, we consider the following two conditions: 
(i) $g(s)$ has no pole at the origin and an odd number of unstable poles (including multiplicities), and 
(ii) $g(s)$ is stabilizable by a static gain and $g(j\omega)$ is real only at $\omega=0$. 
If $g(s)$ satisfies condition (i), then $\rhol_o$ is a lower bound of RIR $\rho_*$, {\it i.e.}, 
\begin{equation} \label{lbo}
\rho_* \geq \rhol_o := 1/|g(0)| \geq \rhol_p
\end{equation}
holds. 
Suppose further that $g(s)$ satisfies condition (ii), 
then the lower bound is tight, {\it i.e.}, $\rho_* = \rhol_o$. 
} 
\end{prop}

{\it Proof:}
The key idea for the proof of the first part is to evaluate the sign of $\Psi(s)/d(s)$ at $s=0$, 
where $\Psi(s) := d(s) - n(s)$ is the characteristic polynomial corresponding to (\ref{eq:che}) 
with a polynomial coprime factorization of $\ell(s) := \delta(s)g(s) = n(s)/d(s)$. 
We have $\Psi(0)/d(0) = 1 - n(0)/d(0) = 1 - \ell(0) < 0$, 
since the Hurwitz stability of $\Psi(s)$ implies $\Psi(0) = d(0) - n(0) > 0$. 
Hence,  $|\delta(0)g(0)| > 1$ which leads to 
$\| \delta \|_{H_\infty} \geq  |\delta(0)|  > 1/| g(0) |$. 
This proves (\ref{lbo}). 
 
Now suppose condition (ii) holds in addition to condition (i) to prove the second part. 
Let $\delta_r\in\IR$ be the perturbation of the smallest magnitude such that 
(a) the closed-loop system has all its poles in the closed left half plane and 
(b) $\delta_r+\eps$ with an arbitrarily small $|\eps|$ can stabilize the closed-loop system:
\[
\delta_r:={\rm arg }\inf_{\delta\in\IR}\{~|\delta|:~ 
1=\delta g(s) ~ \Rightarrow ~ \Re(s)<0~\}.
\]
Note that $\rho_r:=|\delta_r|$ is the real RIR and is an upper bound on the (dynamic) RIR; $\rho_*\leq\rho_r$. 
Now, the characteristic equation $1=\delta_rg(s)$ has a root on the imaginary 
axis. If $s=j\omega$ is a root, then $g(j\omega)=-1/\delta_r$ is a real number. 
Hence, $\omega$ must be zero by condition (ii). It then follows that 
$\rho_*\leq\rho_r=|\delta_r|=|1/g(0)|=\rhol_o$, proving tightness of the lower bound $\rhol_o$. 
\rightline{$\square$}
\subsection{Illustrative Numerical Examples}
\label{subsec:INExamples}

Consider a class of second order feedback systems 
of which the loop transfer function is given by
\begin{equation} \label{h(s)}
h(s) = 2(s-z)/(s^2+s-2) 
\end{equation}
with multiplicative uncertainty, {\it i.e.}, 
\begin{equation} \label{g(s)}
g(s):=\frac{h(s)}{1-h(s)} = \frac{1}{\phi(s) -1}, \hs 
\phi(s) := \frac{1}{h(s)} . 
\end{equation}
Here, $z\in\IR$ is a parameter, the unique real zero of $h(s)$, and 
$h(s)$ is minimum phase (resp. non-minimum phase)  if $z$ is negative (resp. non-negative).  
The characteristic equation for the nominal closed-loop system is given by $1 - h(s)=0$, or $s^2-s+2(z-1)=0$. 
Hence, it is nominally unstable for any $z$.  
This quite simple example turns out to capture various situations 
by changing just one real parameter $z$. 

The robust instability radius and its lower/upper bounds can be calculated analytically as 
follows. 
\footnote{ 
In the context of the multiplicative uncertainty, 
it is reasonable to consider the case where the magnitude of $\delta(s)$ is less than one. 
However, here we consider an arbitrarily large perturbation for illustrative purposes 
to understand the difficulty of the RIR analysis. 
}

\smallskip 

\noindent
{\bf Facts on the RIR:} \;\;
{\em 
Consider $h(s)$ in (\ref{h(s)}) with a parameter $z\in\IR$. 
Define the robust instability radius $\rho_*$ by (\ref{rho}), 
its lower bounds $\rhol_p$ and $\rhol_o$ by (\ref{lub}) and (\ref{lbo}), respectively, 
and its upper bound $\rho_r$ by (\ref{rhostatic}). 
When $0\leq z<1$, we have $\rho_*=\infty$  
due to violation of the PIP condition.  
Otherwise, the result for various cases can be summarized as in Table~\ref{tab:ana}, 
where ``?'' indicates unknown but finite numbers, 
``-'' indicates that there is no $\delta\in\IR$ to stabilize $g(s)$
in the column of $\rho_r$, or $g(s)$ has an even number of unstable pole(s) in the column of $\rhol_o$. 
The lower bounds are determined by minimizing $1/|g(j\omega)|^2$ over $\omega$. 
If the optimizer is $\omega_p = 0$, then $\rhol_p = \sig_0:= 1/|g(0)| = |(z-1)/z|$. 
Otherwise, the minimum $\sig_1$ is achieved at $\omega_p = \omega_1$ where 
\[
\begin{array}{l} 
\omega_1:=\sqrt{\sqrt{\eta(z)}-z^2}, \\
\sig_1 := {\displaystyle\frac{1}{|g(j\omega_1)|}} = \frac{1}{2}\sqrt{2\sqrt{\eta(z)}+5-4c-2z^2}, \\
\eta(z):=(z-1)(z+2)(z^2+3z-2).
\end{array}
\]
The real zeros of 
$\mu(z):=4z^3-z^2-8z+4$  
determine whether the optimizer is zero or non-zero, where 
$b_1=-1.51$, $b_2=0.54$, and $b_3=1.22$ are the roots of $\mu(b_i)=0$.
}

\begin{table}[h]
\centering
\caption{Analytical upper/lower bounds on RIR}
\label{tab:ana} 
\vspace*{-2mm}
\begin{tabular}{|c|c||c||c|c|c|c|} \hline
Case & $z$ & $\rho_*$ & $\rho_r$ & $\rhol_o$ & $\rhol_p$ & $\omega_p$
\\ \hline
(a) & $z \leq b_1$ & $\sig_0$ & $\sig_0$ & $\sig_0$ & $\sig_0$ & $0$ \\ \hline
(b) & $b_1<z<0$ & $\sig_0$ & $\sig_0$ & $\sig_0$ & $\sig_1$ & $\omega_1$ 
\\ \hline\hline
(c) & $1\leq z \leq b_3$ & ? & - & - & $\sig_0$ & $0$  \\ \hline
(d) & $b_3< z \leq 2$ &  ? & - & - & $\sig_1$ & $\omega_1$  \\ \hline
(e) & $2<z$ & ? & $\sig_0$ & - & $\sig_1$ & $\omega_1$  \\ \hline
\end{tabular}
\end{table}

\begin{figure} \centering
\vspace{2mm}
\epsfig{file=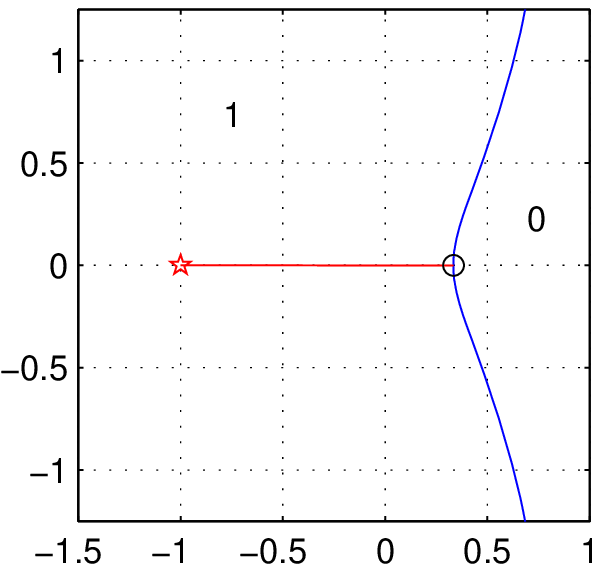,width=24mm} \hspace{2mm}
\epsfig{file=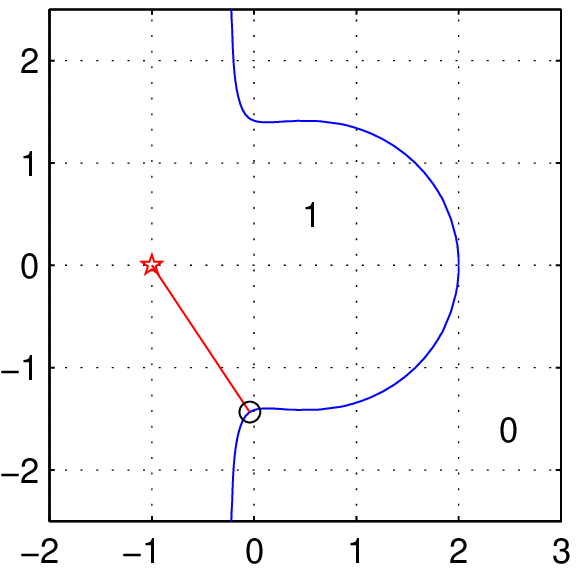,width=23mm} \hspace{2mm}
\epsfig{file=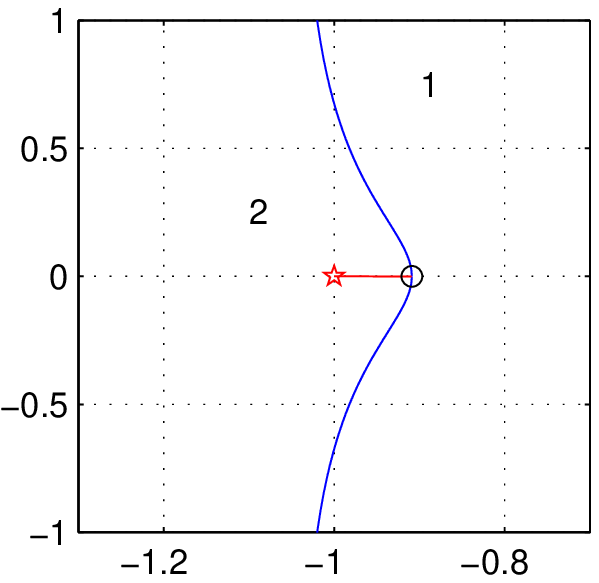,width=24mm} \\
$\mbox{}$ 
\hspace{1mm} (a) $z=-3$
\hspace{7mm} (b) $z=-0.5$
\hspace{7mm} (c) $z=1.1$
\vspace*{3mm} $\mbox{}$ 
\\
\epsfig{file=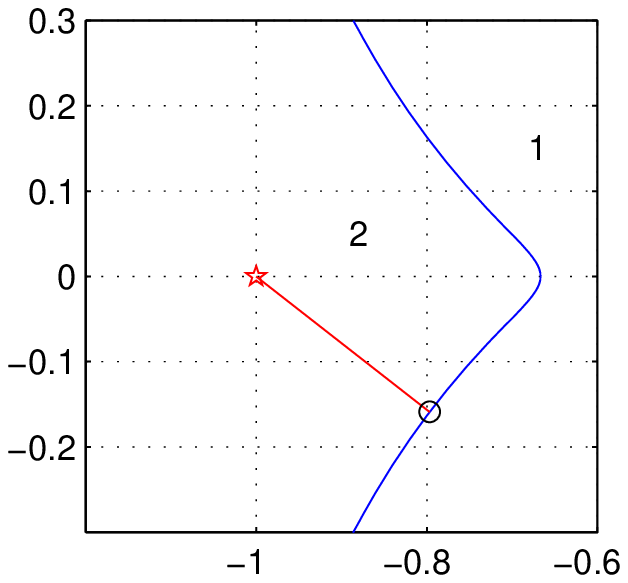,width=25mm} \hspace{2mm}
\epsfig{file=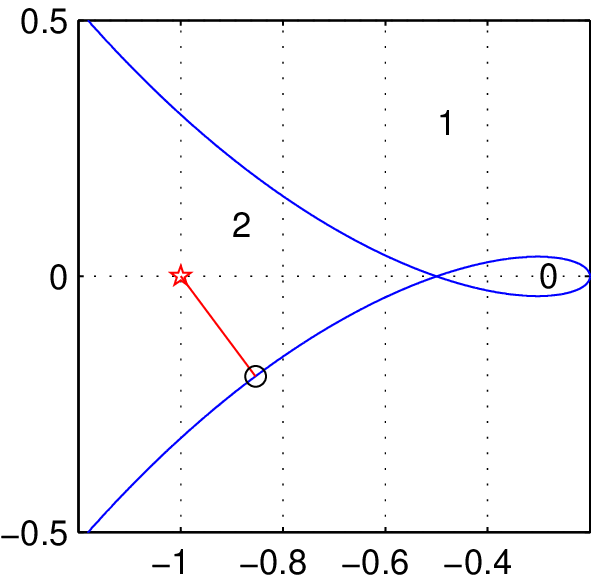,width=24mm} \hspace{2mm} \\
$\mbox{}$ 
\hspace{-5mm} (d) $z=1.5$
\hspace{11mm} (e) $z=5$
\caption{Nyquist plots of $-\phi(s)$ for the five cases in 
Table~\ref{tab:ex}.}
\label{fig:exa}
\vspace*{-3mm}
\end{figure}

We now consider the five cases with different values of $z$ that represent the rows of Table~\ref{tab:ana}.
The numerical result is summarized in Table~\ref{tab:ex}. 
Figure~\ref{fig:exa} shows\footnote{
The Nyquist plots for $-\phi(s)$ instead of $1/g(s)$ is shown 
to conform to the convention where $-1$ is the critical point.
}
the Nyquist plots of $-\phi(s)$, where the number of the characteristic roots 
in the ORHP is indicated in each region divided by the Nyquist plots. 
In each plot, the circle is the projection of the critical point (star) onto the Nyquist 
plot of $- \phi(s)$. The red vector from the circle to the star is 
$\delo := \phi(j\omega_p)-1$, and its length is $\rhol_p$. 

\begin{table}[h]
\caption{Numerical upper/lower bounds on RIR}
\label{tab:ex} 
\centering
\vspace*{-2mm}
\begin{tabular}{|c|c||c||c|c|c|c|} \hline
Case & $z$ & $\rho_*$ & $\rho_r$ & $\rhol_o$ & $\rhol_p$ & $\omega_p$
\\ \hline
(a) & -3   & 1.333 & 1.333  & 1.333 & 1.333 & 0 \\ \hline
(b) & -0.5 & 3 & 3 & 3 & 1.725 & 1.57 \\ \hline\hline
(c) &  1.1 & ?? & - & - & 0.091 & 0  \\ \hline
(d) &  1.5 & ? (0.258) & - & - & 0.258 & 0.80 \\ \hline
(e) &  5   & ? (0.244) & 0.5 & - & 0.244 & 2.76  \\ \hline
\end{tabular} 
\end{table}

We see in case (a) that the projection of $-1$ onto the Nyquist plot $-\phi(j\omega)$ is 
on the boundary of the stability region marked by ``0'' and that $\rhol_p = \rho_r = \rho_o$. 
This means that a slight extension of the projection will stabilize the system, {\it i.e.}, 
$\delta(s) = -(1+\eps)\rhol_p$ with small $\eps>0$ stabilizes, confirming tightness of 
the lower bound (\ref{lub}) in Proposition~\ref{prop:lbub}. 
By this example, we may expect that  
(i) the lower bound is tight when the projection of $-1$ is on the boundary of the stability region 
and that (ii) the lower bound is not tight when the projection is not on the stability boundary. 

It turns out, however, that neither of these is correct as shown by other cases due to a subtle 
difference between the static complex perturbation and the real rational dynamic perturbation.
Actually, case (b) is a counter example of statement (i) in the previous paragraph .
The red vector in Fig.~\ref{fig:exa} (b) clearly starts at one point on the boundary of the 
stability region, which means $\rhol_p = \rho_c = 1.725$, 
but we see from Proposition~\ref{prop:lb0} that $\rho_* = \rhol_o = 3$.  
This is also an example of $\rho_* \neq \rho_c$. 

Three other cases, (c), (d), and (e), 
where the assumptions of Proposition~\ref{prop:lb0} are not satisfied, are related to statement (ii). 
The next section will show for cases (d) and (e) that a stabilizing perturbation 
with slightly larger norm than $\rhol_p$ exists, and hence $\rho_* = \rhol_p$, 
although we were not able to find such a perturbation for case (c).  
This implies that statement (ii) is not always true. 
The situations of the two cases (d) and (e) are slightly different. 
For case (e), we see that $\rhol_p \leq \rho_* \leq \rho_r$. 
However, for case (d), there is no stabilizing static complex perturbation, {\it i.e.}, 
$\rho_c$ ($\leq \rho_r$) is infinite.

\section{Exact RIR Analysis}
\label{sec:ExactAnalysis}
\subsection{Idea: All-Pass Stabilization}
\label{subsec:Idea}

The exact value of the RIR can be found if upper and lower bounds turn out to 
be the same value. The examples in the previous section have shown that the 
lower bounds in Propositions~\ref{prop:lbub} and \ref{prop:lb0} 
are tight for certain classes of $g(s)$.  
However, the upper bound $\rho_r$ is often conservative since the 
class of systems stabilized by a static gain is not large. 
Here we aim to obtain a better upper bound by a dynamic perturbation, 
which will be useful to show tightness of $\varrho_p$ for a class of $g(s)$.  

The lower bound $\varrho_p$ is tight if there is a stabilizing 
$\delta(s)$ of norm arbitrarily close to $\varrho_p$. 
This requires existence of $\delta(s)$, of norm exactly equal to $\varrho_p$,
that marginally stabilizes $g(s)$. Thus we consider the following two-step procedure:
{\bf (Step 1)} marginal stabilization by 
$\delta(s)$ of norm $\|\delta\|_{H_\infty}=\varrho_p$, and 
{\bf (Step 2)} slight modification of $\delta(s)$ to get a stabilizing perturbation. 
A simple calculation shows that the peak gains of $\delta(s)$ and $g(s)$
should occur at the same frequency $\omega_p$, and the closed-loop system
in {\bf Step 1} must have poles at $\pm j\omega_p$. 
While $\omega_p$ is fixed by $g(s)$, the search for $\delta(s)$ in {\bf Step 1}
is complicated in general 
by the constraint $\|\delta\|_{H_\infty}=|\delta(j\omega_p)|$.
To make the analysis tractable, we consider all-pass functions for 
$\delta(s)$ so that the closed-loop pole condition 
$1=\delta(j\omega_p)g(j\omega_p)$ automatically implies satisfaction of
the peak frequency constraint.

\smallskip

\begin{lemma} \label{lemma:AllPass}
{\em 
The positive feedback system consisting of 
\[
\delta(s) = b \cdot \frac{a-s}{a+s}, \hs
g(s) = \frac{\beta_0+\beta_1s+\cdots+\beta_ns^n}
{\alpha_0+\alpha_1s+\cdots+\alpha_ns^n} , \; 
\]
with $\alpha_n \neq 0$, has all the closed-loop poles in the OLHP except for 
a pair of poles at $s=\pm j\omega_c$ if and only if
\begin{equation} \label{eq:Omega_psi}
 \Omega\psi = b N \beta - D \alpha 
\end{equation}
holds for some Hurwitz polynomial
\[
 p(s):=\psi_0 + \psi_1 s + \cdots + \psi_{n-1}s^{n-1},
\]
where $\Omega \in \IR^{(n+2)\times n}$, $D, N \in \IR^{(n+2)\times(n+1)}$, 
$\psi \in \IR^{n}$, and $\alpha, \beta \in \IR^{n+1}$ are defined by
\[
\Omega: = \begin{mat}{c} \omega_c^2I_n \\ 0_{2,n} \end{mat} + \begin{mat}{c} 0_{2,n} \\ I_n \end{mat}, \hs 
\psi := [\psi_0, \cdots, \psi_{n-1}]\t, 
\]
\[
D := \begin{mat}{c} aI_{n+1} \\ 0_{1,n+1} \end{mat} + \begin{mat}{c} 0_{1,n+1} \\ I_{n+1} \end{mat}, \hs 
\alpha := [\alpha_0, \cdots, \alpha_n]\t, 
\]
\[
N := \begin{mat}{c} aI_{n+1} \\ 0_{1,n+1} \end{mat} - \begin{mat}{c} 0_{1,n+1} \\ I_{n+1} \end{mat},  \hs 
\beta := [\beta_0, \cdots, \beta_n]\t. 
\] 
}
\end{lemma}

{\it Proof:}
Denote $g(s)$ by $g(s) = n(s)/d(s)$. 
The closed-loop system has a pair of poles $s=\pm j\omega_c$ and the rest of the poles are 
in the OLHP if and only if $b(a-s)n(s)-(a+s)d(s)=(s^2+\omega_c^2)p(s)$ 
holds for some Hurwitz polynomial $p(s)$. 
Equating the coefficients of the $s^i$ terms, we have the result.
\rightline{$\square$}

Lemma~\ref{lemma:AllPass} with $\omega_c=\omega_p$ provides a computational method 
for checking if {\bf Step 1} is feasible for a given $g(s)$.
First, $\omega_p$ and $\rhol_p$ can be readily obtained by the $L_{\infty}$-norm
computation of $g(s)$. 
Then, for each of the two cases $b=\pm\varrho_p$, we solve the {\em linear} equation 
(\ref{eq:Omega_psi}) for $a$ and $\psi$, and check if $a$ is positive and $p(s)$ is Hurwitz.
If {\bf Step 1} is found feasible, one may check to see if a slight increase of
the gain, $(1+\eps)\delta(s)$ with small $\eps>0$, can strictly stabilize $g(s)$ in {\bf Step 2}. 
 
This method may not work in general, 
but Lemma~\ref{lemma:AllPass} turns out to provide an analytical characterization of 
a class of $g(s)$ for which the method works, as shown in the next subsection. 
The class consists of second order systems but contains the linearized FitzHugh-Nagumo model, 
relevant for the neuronal spike dynamics.

\subsection{Class of Second Order Systems with Exact RIR}
\label{subsec:ExactRIRclass}

The following result provides two classes of second order systems for which
the exact RIR is computable. 

\smallskip

\begin{theorem} \label{theorem:2nd}
{\em
For the second order system represented by 
\begin{equation} \label{g2nd}
g(s)=\frac{rs-1}{s^2+ps+q} , 
\end{equation} 
\begin{itemize}
\item[(i)]
$\rho_*=\varrho_o:=1/|g(0)| = |q|$ holds if 
\begin{equation} \label{c21}
q<0, \hs p+rq > 0 , 
\end{equation}
\item[(ii)]
$\rho_*=\varrho_p:=1/\|g\|_{L_\infty} = 1/|g(j\omega_p)|$ holds if 
\begin{equation} \label{c11}
q>0, \hs p<0, \hs r^2q^2 + 2q - p^2 > 0 ,  
\end{equation}
where $\omega_p^2 = q - p^2/2$ for $r=0$ and otherwise    
\begin{equation}
 \omega_p^2 = [\sqrt{(r^2q^2 + 2q - p^2)r^2+1)} - 1]/r^2 . 
\end{equation}
\end{itemize}
} 
\end{theorem}

See Appendix~\ref{subsec:ProofTheorem} for a proof,
where the results are proven using Proposition~\ref{prop:lb0} for case (i) 
and Proposition~\ref{prop:lbub} with Lemma~\ref{lemma:AllPass} for case (ii).

\smallskip

Let us now confirm the effectiveness of the theorem by using the example 
in the previous section, where 
\[
g(s) = 2z \cdot \big( s/z - 1\big) / \big( s^2 -s +2(z-1)\big) , 
\]
{\it i.e.}, $r=1/z$, $p=-1$, and $q=2(z-1)$.
When $q<0$, statement (i) applies to cases (a) and (b) because $p+rq = -1 + 2(z-1)/z = 1-2/z > 0$. 
Hence, we have $\rho_* = \rhol_p = \rho_r = 1/|g(0)| = 1 - 1/z$. 
When $q>0$, statement (ii) may or may not apply. 
There are two different situations based on 
$\omega_p$, the frequency which gives the maximum gain of $|g(j\omega)|$:
$\omega_p = 0$ for case (c), and $\omega_p > 0$ for cases (d) and (e). 
Since condition (\ref{c11}) holds for cases (d) and (e), 
the first order all-pass function works to stabilize the system, 
and $\rhol_p$ gives the exact RIR $\rho_*$. 
However, we cannot confirm tightness of $\rho_p$ for case (c).

\section{Applications to Neuronal Dynamics}
\label{sec:ApplicationNeuron}

This section is devoted to an application to an analysis of neuronal 
dynamics of excitable membranes for robustly generating action potentials.
We use the FitzHugh-Nagumo (FHN) model \cite{FHNmodel}, which 
is a second order nonlinear system represented by
\begin{equation} \label{FHN}
\begin{array}{l}
c \dot{v} = \psi(v) - w \\
\tau \dot{w} = v+\alpha-\beta w,
\end{array}
\end{equation}
where $c$, $\tau$, $\alpha$, and $\beta$ are positive scalars, and
\[
\psi(v) = v - v^3/3 +i. 
\]
The variable $v(t)$ represents the membrane potential of the neuronal
cell, $w(t)$ is the recovery variable that captures the net effect of
the channel conductances, and $i(t)$ is the current injection input
to the cell, which is assumed constant in the following development.

\begin{figure}[t]
\vspace*{1mm}
\begin{minipage}{0.23\textwidth}
\centering
\epsfig{file=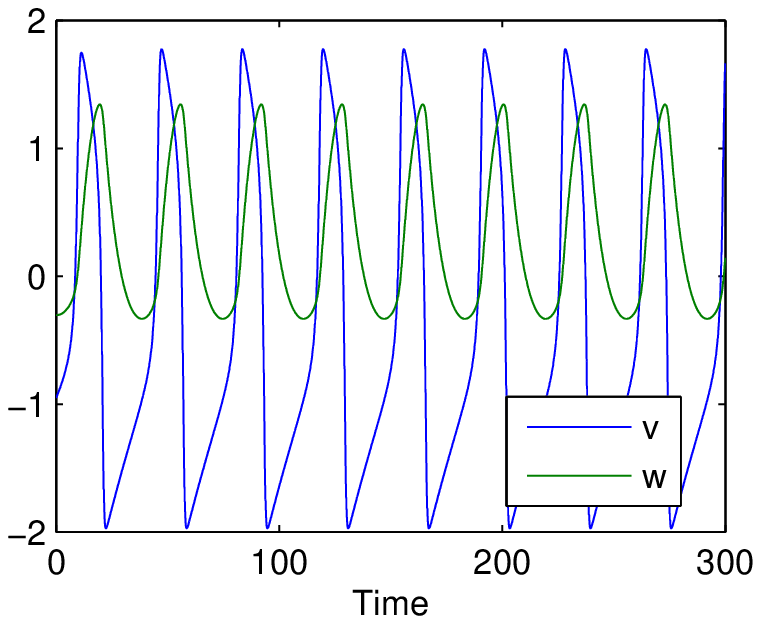,width=33mm}
\vspace*{-2mm}
\caption{FHN model response (nominal, $\delta(s)=0$)}
\label{fig:FHN}
\end{minipage} \hfill
\begin{minipage}{0.23\textwidth}
\centering 
\hspace*{-2mm}
\epsfig{file=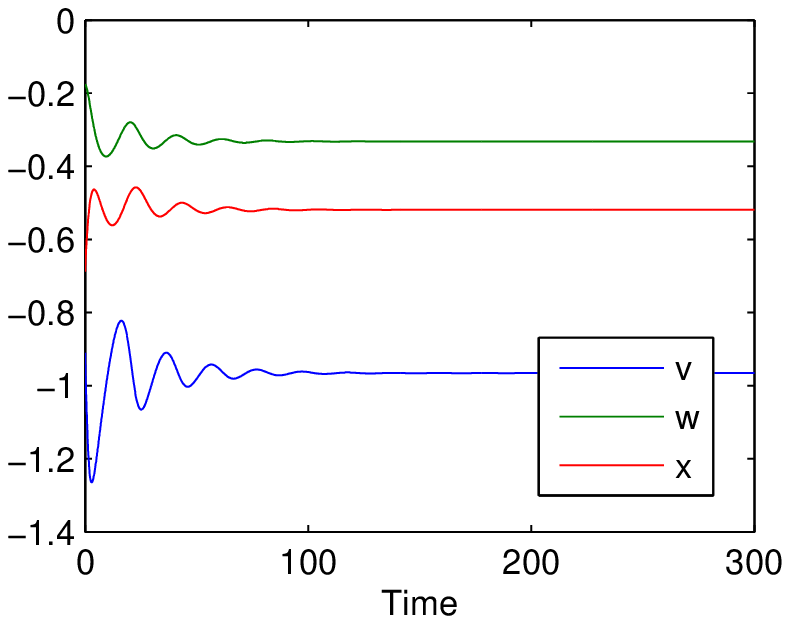,width=33mm}
\vspace*{-2mm}
\caption{FHN model response under a stabilizing $\delta(s)$}
\label{fig:FHNss}
\end{minipage}

\medskip

\begin{minipage}{0.23\textwidth}
\centering
\epsfig{file=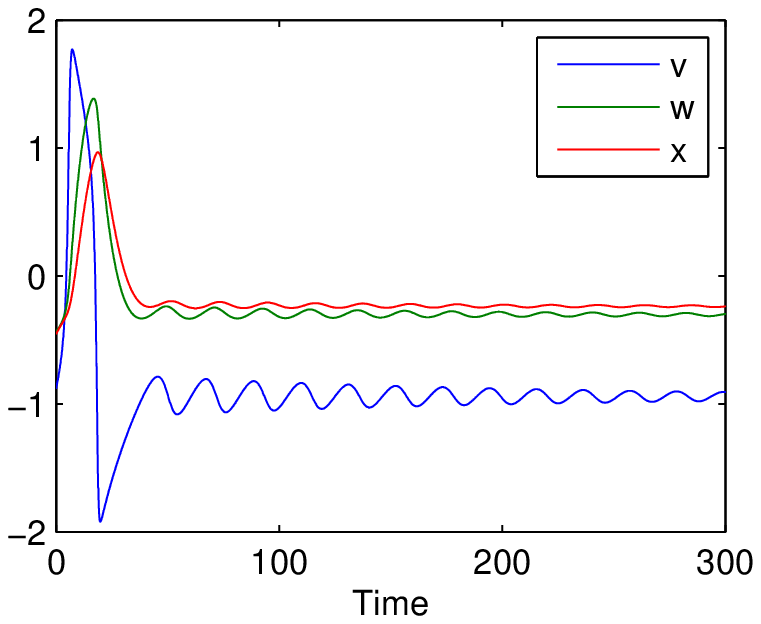,width=33mm}
\vspace*{-2mm}
\caption{FHN response under a stabilizing $\delta(s)$ with $\eps=0.1$}
\label{fig:FHNs}
\end{minipage} \hfill
\begin{minipage}{0.23\textwidth}
\centering
\epsfig{file=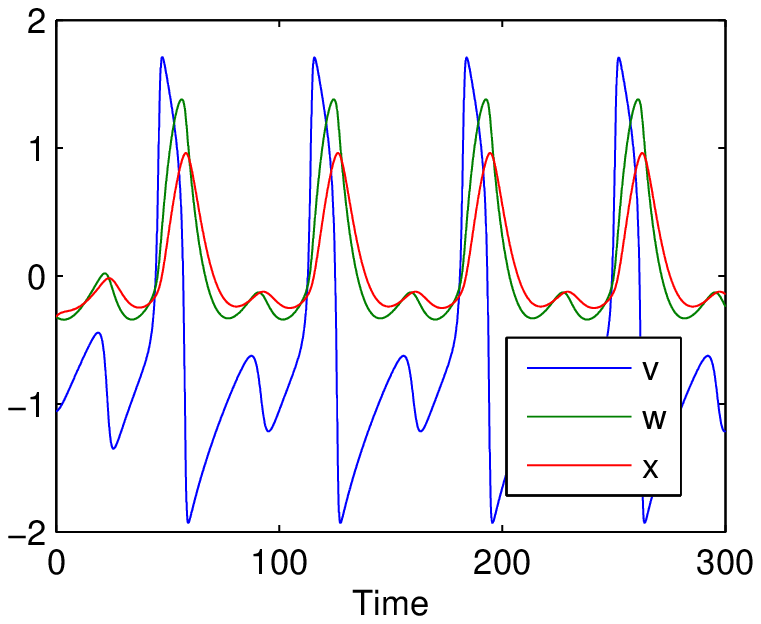,width=33mm}
\vspace*{-2mm}
\caption{FHN response under a destabilizing $\delta(s)$ with $\eps=-0.1$}
\label{fig:FHNu}
\end{minipage}
\vspace*{-2mm}
\end{figure}

We consider the case where $\beta<1$, which guarantees that the system 
has a unique equilibrium point. Based on the shape of the function 
$\psi(v)$, it can be shown that all the trajectories are ultimately bounded. 
Hence, an oscillation occurs when the equilibrium is hyperbolically unstable. 
Typically, the system has a stable limit cycle which is seen as a spike train. 
An example is shown in Fig.~\ref{fig:FHN}, where the parameter values are
\begin{equation} \label{nom}
c=1, \hs \tau=10, \hs \alpha=0.7, \hs \beta=0.8, \hs i=0.4.
\end{equation}
Assuming these values, we will examine robustness of the oscillation against 
unmodeled dynamics.

The FHN model captures the essential dynamics of action 
potential (spike) generation in the simplest way, ignoring the details of 
various channel conductances. Here we model the neglected dynamics by the 
multiplicative uncertainty $\delta(s)w$, where $\delta(s)$ is an uncertain 
stable transfer function. The uncertain FHN model is then given by
\footnote{
With a slight abuse of notation, $\delta(s)w$ means the time-domain signal 
obtained by the inverse Laplace transform of the product of $\delta(s)$ 
and the Laplace transform of $w(t)$.
}
\begin{equation} \label{FHNd}
\begin{array}{l}
c \dot{v} = \psi(v) - (1+\delta(s))w, \\
\tau \dot{w} = v+\alpha-\beta w. 
\end{array}
\end{equation}
Let $(\bar v,\bar w)$ be an equilibrium point, characterized by
\begin{equation} \label{equil}
\psi(\bar v)=(1+e)\bar w, \hs \bar v=\beta \bar w-\alpha,
\end{equation}
where $e:=\delta(0)$. It can be verified that the equilibrium is unique if $1+e>\beta$.  
Linearizing the system around $(\bar{v},\bar{w})$, 
the characteristic equation is given by
\[
1=\delta(s)g_e(s), \; 
g_e(s):=\frac{-1}{c\tau s^2+(\beta c-\tau\gamma)s+1-\beta\gamma}, 
\]
where $\gamma:=\psi'(\bar v)=1-\bar{v}^2$. Note that $g_e(s)$ depends on 
$e$ through $\bar v$, and this dependence is indicated by the subscript. 
For the nominal parameters in (\ref{nom}), the linearization $g_o(s)$ 
({\it i.e.}, $g_e(s)$ with $e=0$) is unstable since $\beta c<\tau\gamma$ holds. 
Thus we have nominal instability. The question is: What is the smallest 
norm of $\delta(s)$ such that $1=\delta(s)g_e(s)$ has all its roots in the 
left half plane?

From Theorem~\ref{theorem:2nd}, the RIR for $g_o(s)$ can be found as
$\rho_*=1/\|g_o\|_{L_\infty}=0.283$.  
This means that the equilibrium remains unstable under perturbations satisfying 
$\|\delta\|_{H_\infty}<\rho_*$, provided the equilibrium does not move by 
the perturbation, {\it i.e.}, $\delta(0)=0$.  
However, there may be perturbation $\delta(s)$ with a nonzero static gain such that 
the equilibrium is moved by $\delta(0)=e$ and becomes stable.  
Indeed, for a stable perturbation $\delta(s)$ of norm $\|\delta\|_{H_\infty}=0.2$, 
the new equilibrium $(\bar v,\bar w)$ in (\ref{equil}) with $e=\delta(0)=-0.2$ is stable, 
and the simulated response converges to $(\bar v,\bar w)$ as shown in 
Fig.~\ref{fig:FHNss}, where $x$ is the state of $\delta(s)$.

A stabilizing perturbation $\delta(s)$ of a smaller norm can be
found as follows. First, solve $|e|=1/\|g_e\|_{L_\infty}$ for $e$ by a line 
search. The solution is found to be $e_o=-0.118$ and the peak gain of 
$g_{e_o}(j\omega)$ is attained at $\omega_p=0.299$. Then by 
Theorem~\ref{theorem:2nd}, the RIR for $g_{e_o}(s)$ is equal to $|e_o|$, 
and a stabilizing perturbation is given by 
$\delta_{\eps}(s) = (1+\eps)\delta_o(s)$ with a sufficiently small $\eps>0$, where 
$\delta_o(s) = 0.118\cdot(s-0.320)/(s+0.320)$ 
is a stable first order all-pass transfer function uniquely determined by
$\delta_o(j\omega_p)=1/g(j\omega_p)$. 
For example, with $\eps=0.1$, the perturbation $\delta_\eps(s)$ has norm 
$\|\delta_\eps\|_{H_\infty}=0.130$, 
and the new equilibrium $(\bar v,\bar w)$ in (\ref{equil}) with 
$e=\delta_eps(0)=-0.130$ is stable.  
Consequently, the simulated response converges to $(\bar v,\bar w)$ as shown in Fig.~\ref{fig:FHNs}.

It turns out that $|e_o|$ is the smallest norm of stabilizing perturbations.
The smallest norm of any stabilizing perturbation $\delta(s)$ with static gain $\delta(0)=e$
is bounded below by the larger of $|e|$ and $\varrho_e:=1/\|g_e\|_{L_\infty}$ because
$\|\delta\|_{H_\infty}\geq|\delta(0)|$ by definition and $\varrho_e$ is a lower bound on the RIR of $g_e(s)$. 
For this particular example, the smallest value of $\max(|e|,\varrho_e)$ over $e$
is achieved at $e=e_o$ where $|e|$ and $\varrho_e$ become coincident. Thus, any perturbation
$\delta(s)$ that stabilizes the equilibrium point must have a norm larger than or equal to $|e_o|$.
Indeed, when $\eps$ for $\delta_\eps(s)$ is negative with a small
$|\eps|$, the perturbation magnitude $\|\delta_\eps\|_{H_\infty}$ is less than
$|e_o|$ and the equilibrium point remains unstable, resulting in a stable
limit cycle as shown in Fig.~\ref{fig:FHNu}.

\section{Conclusion}
\label{sec:Concl}

We have formalized the robust instability problem by introducing a notion of robust instability radius. 
We provided a method for finding the exact RIR for a class of second order systems, 
and the effectiveness has been confirmed by an application to the FitzHugh-Nagumo model. 
Although not reported here, we have also numerically confirmed that 
Theorem 1 should extend to third order systems represented by 
$g(s)=k/((s+\alpha)(s^2 -\beta s + \gamma^2))$ with $0<\beta<\gamma<\alpha$.   
The basic idea in this paper may be applicable for an even broader class of
$g(s)$ that captures a variety of systems appearing in biology 
\cite{HIH:Automatica2020}.

\appendix
\subsection{Proof of Theorem~\ref{theorem:2nd}}
\label{subsec:ProofTheorem}

The proof of (i) for the case of $q<0$ and $\Omega_p=0$ is easy.
The characteristic equation for the closed-loop system with $\delta$ is 
given by $s^2+(p-\delta r)s+q+\delta=0$. 
With $\delta > 1/g(0) = -q$, the two roots are both negative, 
and one can verify that $g(j\omega)$ is real only at $\omega = 0$. 
Proposition~\ref{prop:lb0} then completes the proof of (i).

To prove (ii), let us consider the case of $r=0$ only. 
The proof for the case of $r \neq 0$ is similar albeit with complicated formulas, 
and hence is omitted due to the page limitation.  
To find the critical frequency $\omega_p$, let us define 
$H(\Omega) := 1/| g(j\omega) |^2 = \Omega^2 + (p^2-2q)\Omega + q^2$, 
where $\Omega := \omega^2 \geq 0$. 
Since $dH/d\Omega = 2\Omega - (2q-p^2)$, 
the square of the critical frequency $\Omega_p := \omega_p^2$ is given by 
$\Omega_p = q - p^2/2$ if $2q > p^2$.  
Then, a simple calculation yields $H(\Omega_p) = p^2(4q-p^2)/4 > 0$.

We now apply Lemma~\ref{lemma:AllPass} by setting  $\omega_c^2 = \omega_p^2$ and 
$b = -\sqrt{H(\Omega_p)} = p \sqrt{4q-p^2}/2 < 0$. 
Equation (\ref{eq:Omega_psi}) implies
\begin{eqnarray*}
 x &=& a+p > 0 ,  \hs \omega_p^2 = ap-b+q > 0 , \\
 x\omega_p^2 &=& (a+p)(ap -b +q) = a(q+b) > 0 .
\end{eqnarray*}
These three relations lead to 
\[
 a = (-p + \sqrt{4q -p^2})/2 > 0 , \hs 
 x = (\sqrt{4q -p^2} + p)/2 > 0 ,  
\]
which guarantee the marginal stabilization. 

Next, we show the existence of a stabilizing perturbation by choosing a slightly larger perturbation 
\[
 \delta_\eps(s) = (1+\eps) b \cdot \frac{a-s}{a+s}, \hs \eps > 0 . 
\]
It is readily seen that the corresponding characteristic polynomial is given by 
$s^3 + d_2 s^2 + d_1 s + d_0 = 0$ with 
\begin{eqnarray*}
 d_2 &:=& a + p = (\hat{q} + p)/2 > 0 , \\ 
 d_1 &:=& ap -(1+\eps)b + q = q-p^2/2 - b\eps , \\ 
 d_0 &:=& a\{q + (1+\eps)b\} \nonumber \\ 
 &=& \hat{q}^2(\hat{q} + p)/4 + (\hat{q}- p) b\eps/2 ,    
\end{eqnarray*}
where $\hat{q} := \sqrt{4q-p^2}$. 
We can see that $d_0$ and $d_1$ are both positive for sufficiently small $\eps > 0$, 
and a simple calculation shows that $d_1 d_2 - d_0 = - \eps b \sqrt{4q-p^2} >0$, 
which guarantees the stability of the perturbed feedback system. 
This means that $\rho_* = \rhol_p = |b| = |p| \sqrt{4q-p^2}/2$, 
and hence the proof is completed.

\end{document}